# Observations of Asymmetries in Ionospheric Return Flow During Different Levels of Geomagnetic Activity

J. P. Reistad[1], N. Østgaard[1], K. M. Laundal[1], A. Ohma[1], K. Snekvik[1], P. Tenfjord[1], A. Grocott[2], K. Oksavik[1,3,4], S. E. Milan[1,5], and S. Haaland[1,6]

[1]Birkeland Centre for Space Science, University of Bergen, Bergen, Norway, [2]Physics Department, Lancaster University, Lancaster, UK, [3]Arctic Geophysics, University Centre in Svalbard, Longyearbyen, Norway, [4]Center for Space Science and Engineering Research, Virginia Tech, Blacksburg, VA, USA, [5]Department of Physics and Astronomy, University of Leicester, Leicester, UK, [6]Max-Planck Institute for Solar Systems Research, Göttingen, Germany

**Abstract** It is known that the magnetic field of the Earth's closed magnetosphere can be highly displaced from the quiet-day configuration when interacting with the interplanetary magnetic field (IMF), an asymmetry largely controlled by the dawn-dusk component of the IMF. The corresponding ionospheric convection has revealed that footprints in one hemisphere tend to move faster to reduce the displacement, a process we refer to as the restoring of symmetry. Although the influence on the return flow convection from the process of restoring symmetry has been shown to be strongly controlled by the IMF, the influence from internal magnetospheric processes has been less investigated. We use 14 years of line-of-sight measurements of the ionospheric plasma convection from the Super Dual Auroral Radar Network to produce high-latitude convection maps sorted by season, IMF, and geomagnetic activity. We find that the restoring symmetry flows dominate the average convection pattern in the nightside ionosphere during low levels of magnetotail activity. For increasing magnetotail activity, signatures of the restoring symmetry process become less and less pronounced in the global average convection maps. We suggest that tail reconnection acts to reduce the asymmetric state of the closed magnetosphere by removing the asymmetric pressure distribution in the tail set up by the IMF $B_y$ interaction. During active periods the nightside magnetosphere will therefore reach a more symmetric configuration on a global scale. These results are relevant for better understanding the dynamics of flux tubes in the asymmetric geospace, which is the most common state of the system.

**Plain Language Summary** In this study we use observations of plasma drift from the Earth's ionosphere to study the symmetry of the Earth's magnetosphere on a large scale. On this global scale we say that the magnetic field is asymmetric when the field lines connecting the two hemispheres are displaced from their usual location. This can happen when the magnetosphere interact with the interplanetary magnetic field, especially when the latter has a significant magnitude in the east-west direction. The major discovery of this study is that geomagnetic activity related to processes within the magnetosphere (tail reconnection) also seem to influence the degree of global asymmetry in the system. We find that the magnetosphere can become very asymmetric during periods of low geomagnetic activity, while it is more symmetric during times with higher activity. These results give us a better understanding of the processes leading to an asymmetric magnetosphere, which is needed to better understand the complex near-Earth space system that is becoming increasingly important for our society.

## 1. Introduction

The Earth's magnetospheric response to solar wind forcing can to a large extent be explained as a two-stage process. The two main stages are (1) opening and (2) closure of magnetic flux through dayside and tail reconnection, respectively, a description usually referred to as the Expanding/Contracting Polar Cap paradigm (Cowley & Lockwood, 1992). Despite knowing that the combination of the two stages controls most of the large-scale dynamics, it is often convenient to separate effects and look at influence on the system by one of the two in order to gain detailed insight. One such example is the numerous studies of interplanetary magnetic field (IMF) clock angle influence on various high-latitude large-scale electrodynamic properties such as electric and magnetic fields, currents, convection, conductivity, etc. (e.g., Haaland et al., 2007; Pettigrew et al., 2010; Weimer, 2005; Weimer et al., 2010). Such studies mostly agree on the large-scale features of the global distributions of these parameters in response to IMF clock angle. However, their main caveat is that they are large statistical averages, often sorted only by the IMF orientation and magnitude. Hence, they mix times when only Stage 1 dominates (i.e., largely controlled by the IMF clock angle), and times when also significant nightside reconnection is present, which drastically changes the state of the system.

In this paper we investigate the asymmetric state of the closed nightside magnetosphere, as influenced by both external (mainly by IMF $B_y$) and internal processes, with focus on tail reconnection. In the following we briefly explain how the IMF interacting with the magnetosphere can lead to an asymmetric closed magnetosphere: Earlier observations have shown that the closed magnetic field lines from the inner edge of the plasma sheet, and further tailward, respond to IMF $B_y$ by inducing a local $B_y$ in the same direction (Cowley & Hughes, 1983; Lui, 1984; Petrukovich, 2011; Wing et al., 1995). A description of how this situation could arise was put forward by Cowley (1981) and Cowley and Lockwood (1992). Based on this we now understand the appearance of a local $B_y$ in the closed magnetosphere as a result of the asymmetric loading and/or rearranging of magnetic flux in the two magnetotail lobes. Here loading refers to dayside reconnection opening magnetic flux, while rearranging of open magnetic flux will be a result of lobe reconnection, usually considered to be most important during northward IMF. As described by Cowley (1981) and Cowley and Lockwood (1992), the loading/rearranging of magnetic flux becomes asymmetric between hemispheres in the presence of IMF $B_y \neq 0$ due to the tension forces acting on the newly reconnected field lines. In this description, $B_y$ penetrates in with the convection and can therefore not appear on closed field lines before such asymmetric open field lines have reconnected in the tail. More recently, Khurana et al. (1996) and Tenfjord et al. (2015) introduced a new variant of the description of how $B_y$ on closed field lines can develop. In this description, the initial result of asymmetric loading and/or rearranging of magnetic flux is an asymmetric magnetic pressure distribution across the magnetotail, established on a short timescale compared to the time $B_y$ will "penetrate" in by convection through the lobes in the Cowley (1981) and Cowley and Lockwood (1992) description. As a result, magnetic pressure gradient forces will initiate convection acting to equilibrate the asymmetry in the magnetic pressure distribution. This process can affect also the closed field lines, without invoking tail reconnection. The opposite directed pressure gradient forces in the northern and southern part of the closed field lines is what can lead to a shear flow that will *induce* a $B_y$ component in the closed magnetosphere. Superposed epoch studies of the $B_y$ response at geosynchronous orbit to IMF $B_y$ reveal a prompt response for both northward and southward IMF (Tenfjord et al., 2017, 2018), as expected from this mechanism. On the other hand, the study by Browett et al. (2017) suggests that in the more distant tail (14–19 $R_E$), a longer response time can also exist, more consistent with the Cowley (1981) and Cowley and Lockwood (1992) explanation of how $B_y$ changes in the magnetosphere.

According to Tenfjord et al. (2015), the stresses inducing the $B_y$ in the magnetosphere will be transmitted toward the ionosphere and act to displace the footprints of the field line in opposite directions (east-west) in the two hemispheres. Studies using simultaneous imaging of the global aurora in both hemispheres have frequently observed such a displacement of auroral features that is inferred to be conjugate (on the same field line; e.g., Østgaard et al., 2004, 2011; Reistad et al., 2013, 2016). Such observations show an overall agreement with the present IMF $B_y$ when it comes to the direction of the displacement of conjugate regions. However, there is much variability in the displacement during events when the IMF is fairly stable (Østgaard et al., 2011; Reistad et al., 2016), pointing toward other important controlling factors.

Grocott et al. (2004) were the first to relate observations of fast (∼ 1, 000 m/s) azimuthal (east-west) plasma flows in the nightside ionosphere auroral zone to the dynamic evolution of a closed field line being highly displaced due to IMF $B_y$. The direction of the ionospheric azimuthal fast flows was found to depend on IMF $B_y$, and be oppositely directed in the two hemispheres, in a manner that the asymmetry in footprint location between hemispheres would be reduced. Hence, the Northern Hemisphere fast flows were eastward for positive IMF $B_y$ (dawn cell) and westward for negative IMF $B_y$ (dusk cell). Furthermore, Grocott et al. (2007) and Pitkänen et al. (2015) found conjugate in situ signatures of the azimuthal flows in the magnetosphere, pointing toward a large-scale reconfiguration process. From further investigations (Grocott et al., 2005) and earlier work by Nishida et al. (1998), Grocott et al. (2007) proposed a mechanism for the flux transport associated with these azimuthal fast flows. Grocott et al. (2007) explained the observations in terms of a large-scale reconfiguration of an asymmetric tail, where the azimuthal fast flows act to remove the asymmetry caused by IMF $B_y$. They suggest that the observed azimuthal fast flows are the ionospheric manifestation of this large-scale reconfiguration, occurring when field lines reconnect in the distant tail with asymmetric footprints due to the IMF $B_y$ influence.

The ionospheric azimuthal fast flows are frequently termed TRINNI (Tail Reconnection during IMF-Northward, Non-substorm Intervals; Milan et al., 2005) as they have a strong preference to be observed during northward but $B_y$-dominating conditions (Grocott et al., 2008) and are explained as a result of tail reconnection. The mechanism proposed by Grocott et al. (2007), however, does not provide an obvious explanation for why signatures of this large-scale reconfiguration are not as frequently seen during southward IMF conditions. Hence, the present understanding of what control the degree of asymmetry in the system is not satisfactory.

The aim of the present paper is to enhance the understanding of what determines the degree of asymmetry in the closed nightside magnetosphere, with emphasis on the influence from tail activity. Our approach is to produce average maps of the high-latitude ionospheric plasma convection during certain intervals of external and internal driving. Although the system is highly dynamical in response to these two types of driving, it is assumed that the system reaches a steady state during the specific intervals of internal and external driving, so that our average maps of convection are representative for the system during these conditions. We examine the influence on the nightside convection pattern from tail activity and show that the influence from external asymmetric forcing becomes less and less evident for increasing tail activity; that is, the nightside magnetosphere becomes more symmetric during increased tail activity.

The following section describes the methodology used to derive the ionospheric convection maps and the data selection. Section 3 presents the results, followed by a discussion and conclusion in sections 4 and 5, respectively.

## 2. Method

In this study we produce maps of average ionospheric plasma convection at high latitudes during various conditions related to IMF, solar illumination, and geomagnetic activity to be analyzed with respect to external and internal forcing of the system. The different sources of data as well as details on how the data were processed are described in the following subsections.

### 2.1. IMF Data

This study will investigate how the IMF clock angle, among other parameters, influences the global ionospheric convection on closed field lines. Hence, we want to only use convection data from times when the IMF is stable. This is achieved by using the method of bias filtering (Haaland et al., 2007) of the IMF data, where stability is quantified as the length of the bias vector. The bias vector is found from adding unit vectors along the individual IMF observation (in $yz$ plane) and normalizing to the number of observations over the period in which stability is being considered. For stable IMF, the bias vector length will be 1, while randomly distributed IMF orientations within the filter window would have resulted in bias filter values close to 0. We use the 1 min OMNI data product from National Aeronautics and Space Administration's Space Physics Data Facility (King & Papitashvili, 2005), represented in the Geocentric Solar Magnetic (GSM) coordinate system. These Solar Wind (SW) and IMF data are time shifted to the nose of the Earth's bow shock. When calculating the length of the IMF bias vector we use a 2-hr rolling interval. This is longer than the reconfiguration time of the magnetic field at geosynchronous orbit, found to be $\sim$ 50 min by Tenfjord et al. (2017, 2018). The 2-hr interval is selected to be before the measurement and also includes 10 min ahead in time to remove influence from a possible uncertainty in the time shift from the L1 point. We use the same threshold for bias vector length as Haaland et al. (2007), 0.96, which is satisfied 27% of the time when using the 2-hr rolling window approach. During this interval, a linear variation in the IMF clock angle of 56° gives a bias vector length of 0.96. Furthermore, we require that IMF data are available in two thirds of the 2-hr interval. The IMF clock angle, $\theta$, is then determined from the mean IMF $B_y$ and IMF $B_z$ in the same 2-hr interval and defined as the angle of the IMF$_{yz}$ vector from the GSM $z$ axis, with positive sign for positive IMF $B_y$.

### 2.2. Dipole Tilt Angle

We define the dipole tilt angle as the angle between the centered magnetic dipole axis and the GSM $z$ axis, in the GSM $xz$ plane. By convention, positive values for the dipole tilt angle correspond to northern summer. The dipole tilt angle will to a large extent quantify the amount of sunlight received by each hemisphere. As a single number, the dipole tilt angle is a more accurate measure of the hemispheric asymmetry in insolation compared to day of year. However, as pointed out by Laundal et al. (2016), significant variations in solar illumination of the two polar caps are possible for the same dipole tilt angle. This is due to structures in the Earth's magnetic field, different from the simple dipole, making a realistic magnetic coordinate system distorted.

### 2.3. Measure of Geomagnetic Activity

To sort our data according to geomagnetic activity, we use the AL index provided by the World Data Centre for Geomagnetism, Kyoto. This is a network of 12 magnetometers evenly spread around the Northern Hemisphere auroral zone. The AL index is the minimum value of the *H* component of the perturbation field at any time. It is used as a quantification of the level of geomagnetic disturbance, often related to processes in the nightside magnetosphere.

### 2.4. Global Ionospheric Convection Maps

To produce the global ionospheric convection maps based on the conditions discussed above, we use data from the Super Dual Auroral Radar Network (SuperDARN; Chisham et al., 2007; Greenwald et al., 1995). SuperDARN radars operate by transmitting high-frequency radio signals that refract in the ionosphere and backscatter from decameter-scale, magnetic field-aligned irregularities in the electron density. The backscattered signal, mainly from the *F* region ionosphere, experiences a Doppler shift that is proportional to the line-of-sight (LOS) component of the plasma drift velocity. In the following, the LOS velocity component is always presented as a positive value, having an associated azimuth angle to represent the LOS direction. The LOS measurements have been boxcar filtered and gridded on an equal area grid (Ruohoniemi & Baker, 1998). For every gridded LOS measurement we further require that the standard deviation of the individual LOS measurements prior to the mentioned filtering process is less than 40% of the median of the same distribution. This is to reduce the uncertainty of the LOS observations.

In this paper we use data from all available SuperDARN radars in the Northern Hemisphere in the time period 2000–2013. We calculate the binned average of the selected LOS measurements satisfying the IMF, tilt angle, and AL criteria. The binned average is calculated in each equal area grid cell as defined by magnetic local time (MLT) and magnetic latitude (MLAT) coordinates, where the latitudinal length is 2° and the MLT width is 6 hr in the most poleward bins (87°–89° MLAT). The average plasma convection vector in each bin is determined by a least squares fit to all the LOS measurements within each bin. This is similar to the method described by Förster et al. (2008), who derived global maps of neutral wind data from cross-track in situ observations of the neutral wind at ∼400-km altitude. Our validating metrics of the fit as described below are largely following their methodology.

An example of this procedure is shown in Figure 1. Here the grid cell where MLAT ∈ [67°, 69°] and MLT ∈ [21, 22] hr is shown for Northern Hemisphere winter (dipole tilt < −10°), IMF clock angle $\theta \in [45°, 90°]$ corresponding to positive IMF $B_y$ and northward IMF $B_z$, and during quiet geomagnetic conditions (AL > −50 nT). This is the same conditions as will be presented in Figure 3B. The observed LOS velocity is plotted as a function of the LOS azimuthal direction at echo location, where 0° is along the magnetic meridian (northward) and positive to the east. From Figure 1 it is evident that the part of the distribution below ∼ 100 m/s is missing. This is because the ground is often reflecting a signal to the radar, and the processing scheme making the gridex files remove small velocities to eliminate this problem. However, this also removes actual observations from the *F* region of small velocities.

In our fitting of the observations to an average convection vector we seek a solution that minimizes $\delta f$ in

$$\delta f = \sum_{i=1}^{n} \left( Vlos_i - \vec{k}_i \cdot \vec{V} \right)^2 \tag{1}$$

Here $Vlos_i$ and $\vec{k}_i$ are the LOS velocity and the unit vector along the LOS direction at echo location of observation *i*, respectively. $\vec{V}$ is our average plasma velocity vector in the respective bin that minimizes $\delta f$. $\vec{V}$ projected onto the azimuthal directions ∈ [−180, 180] is shown as the red line in Figure 1, and the azimuth angle corresponding to $\vec{V}$ is indicated with the red dot. As a measure of the spread from this line one can use the standard deviation from the red line, $\sigma$, defined as (Förster et al., 2008)

$$\sigma = \sqrt{\frac{1}{n-1} \sum_{i=1}^{n} \left( Vlos_i - \vec{k}_i \cdot \vec{V} \right)^2} \tag{2}$$

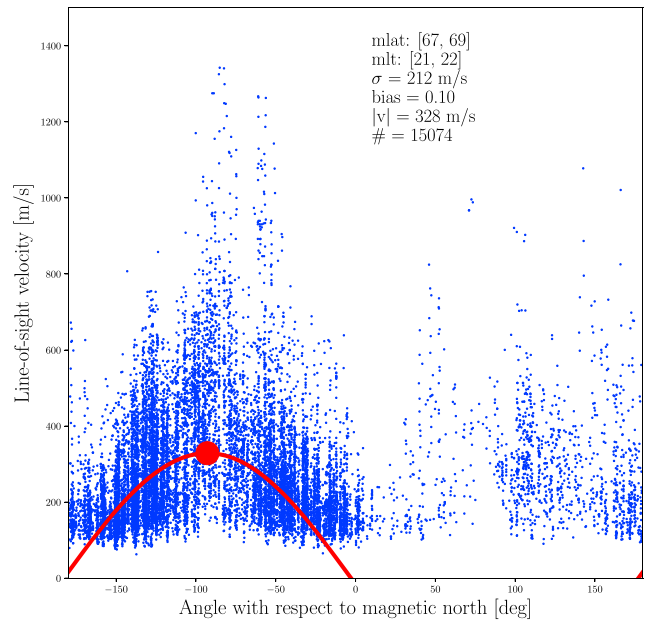

**Figure 1.** Example of the least squares fitting process in an MLT/MLAT grid cell inside the dusk cell return flow. The selection parameters are the same as in Figure 3b. The observed LOS velocities (*y* axis) are shown as a function of the azimuth angle at echo location (*x* axis) as blue dots. The mean plasma velocity vector within this bin, $\vec{V}$, is determined by the method of least squares and is indicated by the red dot. The projection of $\vec{V}$ onto the azimuthal directions is shown as the red line. MLT = magnetic local time; MLAT = magnetic latitude; LOS = line of sight.

In the example in Figure 1, the standard deviation is 212 m/s. Due to the fixed locations of the radar beams in geographic coordinates, not all azimuth directions will be sampled equally frequently. To quantify how well the various directions are sampled within an MLT/MLAT grid cell, we use a bias vector approach, similar to the one used for the IMF stability. The radar measurements are different from the full vector measurements of the IMF when applying the bias filtering technique as the radars are only able to resolve if the convection is toward or away from the radar. To take this into account we do the following before calculating the bias vector for the look directions: Negative azimuth values are shifted 180° so each observation is allocated the same azimuth regardless if observation is toward or away from the radar. The azimuth distribution is then expanded to a full circle again by multiplying every azimuth by 2. The bias vector is then found by adding all the new unit vectors and divide each component (north, east) by the number of observations. If all directions are represented equally, the length becomes zero. For an increasing bias in look direction, the length of the bias vector approaches unity. For the example in Figure 1, the bias vector length is 0.10 indicating a very good coverage of the different azimuths. Similar to Förster et al. (2008), we require the length of the bias vector to be < 0.9 to consider the grid cell in the further analysis. A bias vector length of 0.9 corresponds to the bias in look direction if the half-circle distribution of observed azimuth directions follows a normal distribution with full width at half maximum of 31°.

Figure 2a shows the result of the fit in every grid cell > 60° MLAT in the Northern Hemisphere during the conditions used in Figure 1 (local winter, IMF clock angles between 45° and 90°, and AL > −50 nT). Only grid cells having > 100 observations and bias < 0.9 are shown in this panel and are used in the further analysis. With these limits we avoid showing grid cells with a solution that is intuitively unrealistic. The estimated average convection is indicated both with a black vector pin and with color reflecting the total magnitude. Figure 2b shows the corresponding number of LOS measurements in each grid cell on a logarithmic scale, Figure 2c shows the standard deviation discussed above, and Figure 2d shows the length of the bias vector. We do not have any criterion for the standard deviation. However, the standard deviation is a good indicator of regions of variability in the convection pattern. Most notably we have large variations in the region close to the typical location of the boundary between open and closed field lines, especially on the banana-shaped cell, and the region in the nightside that sometimes can be a part of either the dawn cell or the dusk cell. To select the return flow region we therefore focus on grid cells equatorward of the region of elevated standard deviation, which is most evident in the dawn region in Figure 2c.

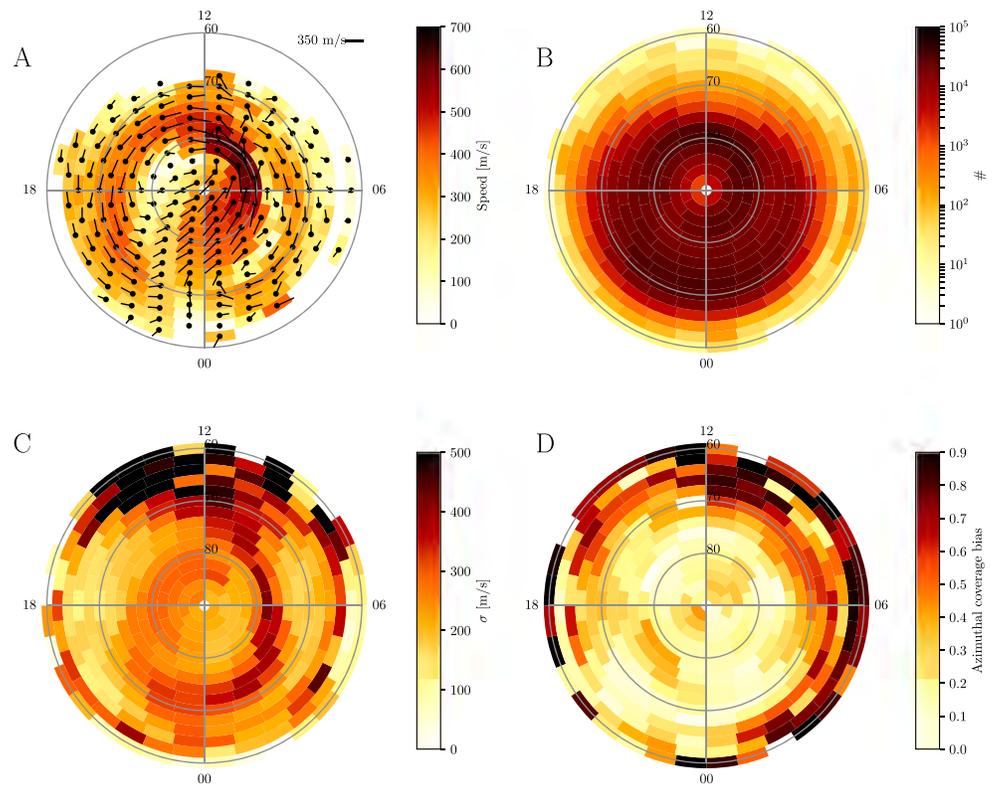

**Figure 2.** (a) Example of the resulting global ionospheric convection map from the fitting described in section 2.4. (b) Number of individual LOS observations in each grid cell. (c) Standard deviation as described in equation (2) for every grid cell. (d) Directional bias in observing direction. Selection parameters are the same as used in Figures 1 and 3b. LOS = line of sight.

## 3. Results

As our focus is on the effect of various levels of tail activity on the convection of closed field lines, we only present the nightside part of the global convection maps. We have selected four different intervals of the IMF clock angle, $\theta$: northward and $B_y$ dominated ($\theta \in [-90°, -45°]$ and $\theta \in [45°, 90°]$), and southward and $B_y$ dominated ($\theta \in [-135°, -90°]$ and $\theta \in [90°, 135°]$). We distinguish between northward and southward IMF since earlier studies have indicated a difference in occurrence of ionospheric azimuthal fast flows between northward and southward IMF (Grocott et al., 2008). Also the energy input from the SW is different in the two cases. For the different levels of tail activity we use three intervals: quiet levels (AL > −50 nT), moderate activity (−150 nT < AL < −50 nT), and active periods (AL < −150 nT). The data are further divided into three categories depending on season: winter (tilt < −10°), equinox (|tilt| < 10°), and summer (tilt > 10°). We only present data from the Northern Hemisphere.

### 3.1. Northern Hemisphere Winter

We start by showing the winter plots in Figure 3. Here we see the average nightside ionospheric convection pattern for the 12 different IMF clock angle ($\theta$) and AL combinations indicated by Figure 3 top row and left column. The panels are organized to ease the comparison of the same |$\theta$| (e.g., comparing Figures 3a and 3b, and 3c and 3d). In addition to the pins representing the average convection vector in each bin, the total convection speed is also indicated by color. A two cell pattern in the nightside is seen in all panels, as earlier shown by numerous studies (e.g., Haaland et al., 2007; Pettigrew et al., 2010; Weimer, 2005).

We want to compare how the speed of the nightside convection on closed field lines in the dawn and dusk convection cells varies for different IMF directions ($\theta$) and tail activity. Before comparing the convection speed in the two cells we need to determine the average speed in each cell separately. In doing so, we first identify the nightside MLT location where the convection splits into a dawn cell and a dusk cell, the nightside convection throat. This location is determined from where the average eastward velocity component between 60° and 70° MLAT goes from positive to negative. We highlight this location with a thick blue line in every panel in Figure 3. We then quantify the return flow speed in the dusk and dawn convection cells by calculating the mean of the convection velocity in every grid cell between 60° and 70° MLAT in a 6-hr wide MLT sector toward the west or east of the blue line, respectively. In each panel in Figure 3 at 21 and 03 MLT we print the estimated dusk and dawn convection speed, respectively, in units of meter per second. We expect the process of restoring symmetry to affect the dawn and dusk convection speed differently, enhancing the flow speed in one cell while decreasing the flow speed of the other cell. We therefore use the dawn/dusk ratio as a metric of the influence from the restoring of symmetry process, and this value is printed at the bottom of each panel in Figure 3. We use this metric when analyzing how IMF and tail activity affect the relative convection speed in the dawn and dusk cells, rather than pointing to the individual maps. In this way it will be easier to point out the main results from this study. In our analysis we focus on the trends of this ratio rather than the value itself, which to some extent will reduce the impact of possible biases introduced when quantifying the convection speed, as, for example, correction for the corotation of the radars in the MLT/MLAT frame (we use velocities relative to the ground). Taking corotation into account would increase the value of the ratio. However, the main trends of the ratio still remain. In the following, we will use the term asymmetric when the dawn/dusk ratio during the same AL interval and season is different for one direction of $B_y$ compared to the other (e.g., comparing the ratio in Figures 3a and 3b). The fact that the ratio itself is different from 1 is not the focus of this study but rather how this ratio changes with geomagnetic activity, and if there is a different trend for negative and positive IMF clock angle.

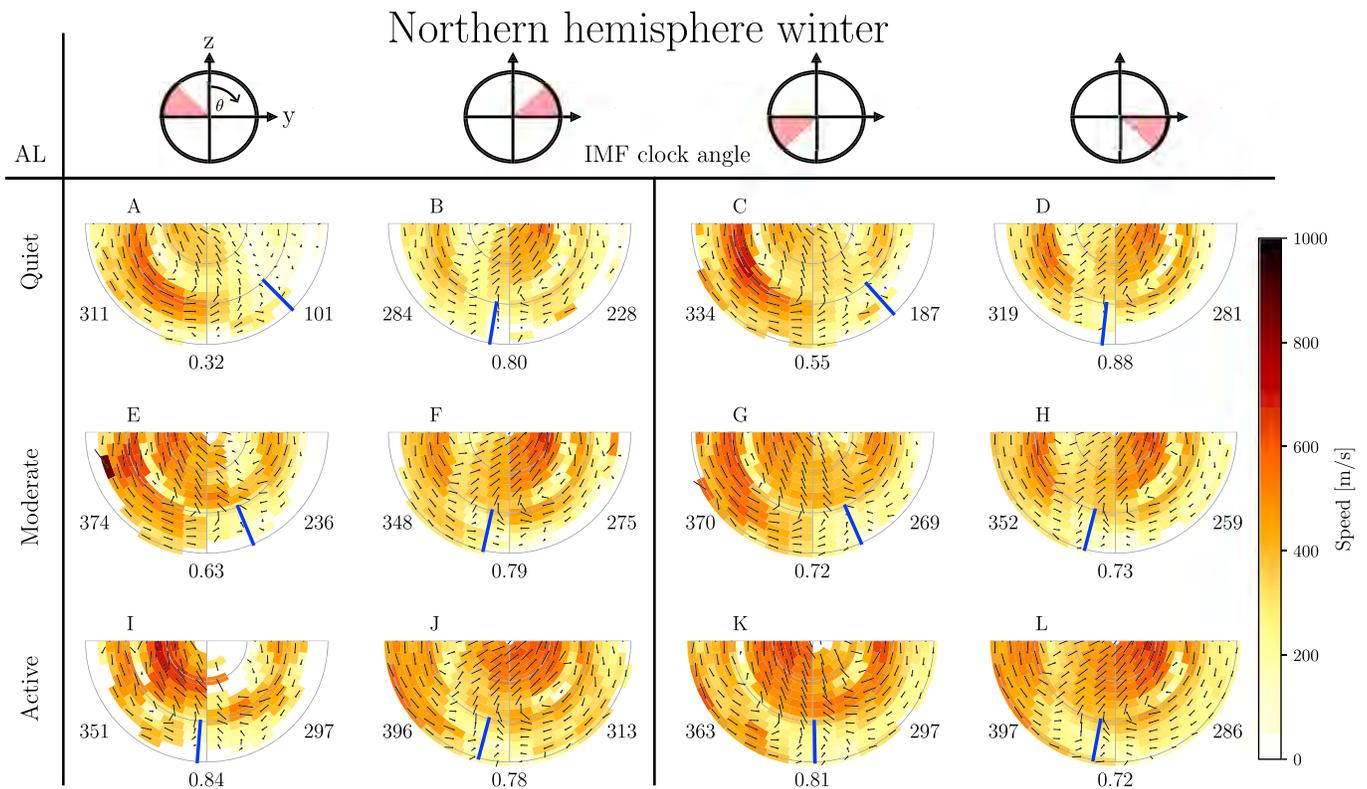

**Figure 3.** Maps of nightside ionospheric convection for winter conditions in the Northern Hemisphere (dipole tilt < −10°). Black pins indicate the estimated average convection vector based on SuperDARN LOS measurements, and its magnitude is also indicated with color. The measurements are selected for the IMF clock angle intervals indicated at the top row, and for the geomagnetic activity levels indicated to the left: quiet (AL > −50 nT), moderate (−150 nT < AL < −50 nT), and active (AL < −150 nT). Blue bold lines indicate the transition region between the dusk and the dawn convection cells. Average speeds of the return flow in the two cells are printed in each panel in the respective MLT sector in units of meter per second. The ratio of the dawn/dusk convection speed is printed at the bottom center of each panel. SuperDARN = Super Dual Auroral Radar Network; LOS = line of sight; IMF = interplanetary magnetic field; MLT = magnetic local time.

Comparing pairs of positive and negative IMF $B_y$ for both northward and southward IMF, during the same AL interval (e.g., Figures 3a and 3b, and 3c and 3d), we see that the dawn/dusk convection speed ratio is smaller for negative IMF $B_y$ compared to the corresponding positive IMF $B_y$ panel during the quiet and moderate AL activity bins. This is consistent with negative IMF $B_y$ enhancing dusk cell convection and/or positive IMF $B_y$ enhancing dawnside convection. We also observe that during negative IMF clock angles, the ratio increases with increasing |AL|, and during positive IMF clock angles the ratio decreases with increasing |AL|, with no exception. Regardless of the initial value of the ratio, for both northward and southward IMF, the asymmetry between the ratios (positive and negative $\theta$) is decreased during the most active conditions compared to the most quiet conditions. This is our main point from Figure 3. The decrease of asymmetry between the same $|\theta|$ interval with increasing |AL| is clearly seen for northward IMF but is also seen during southward IMF. In both cases the asymmetry is actually reversed in the highest |AL| interval.

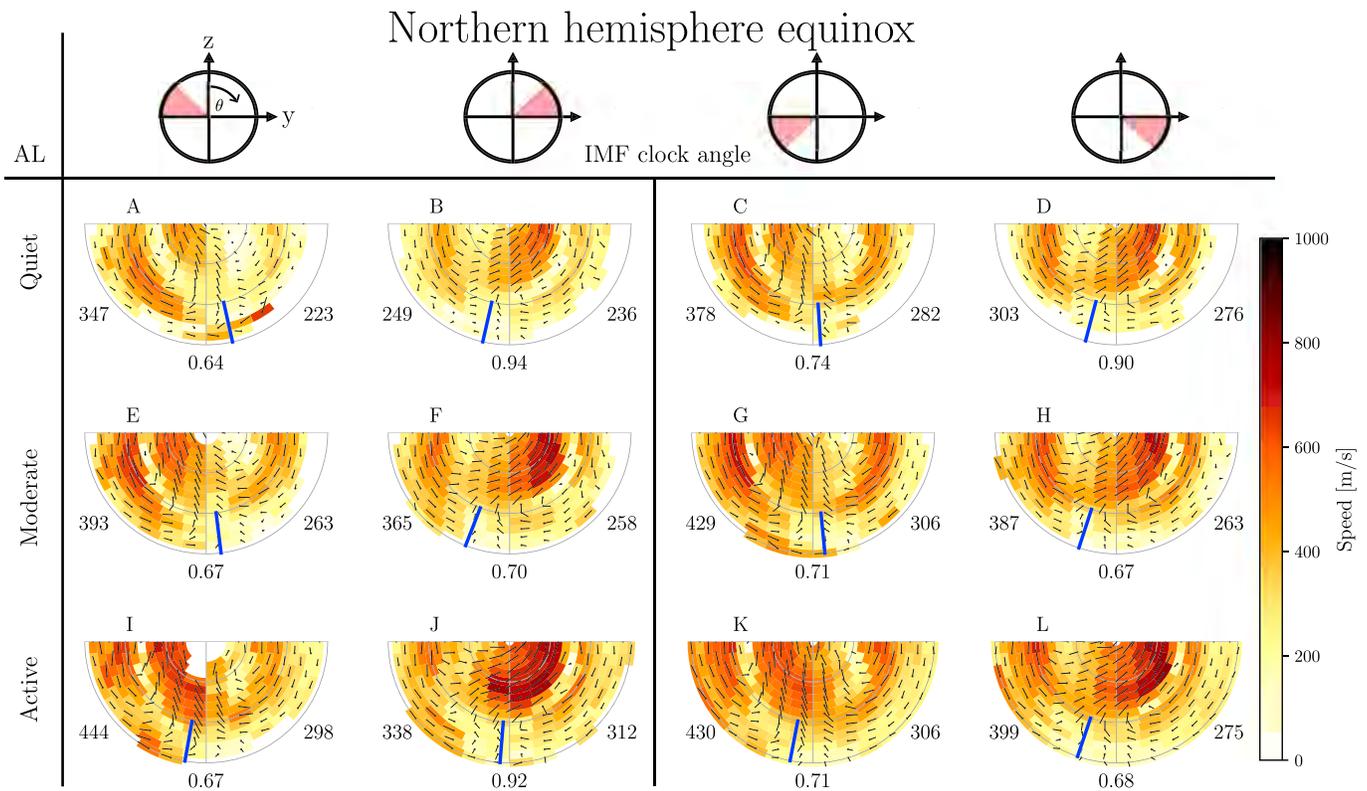

**Figure 4.** Maps of nightside ionospheric convection during equinox conditions in the Northern Hemisphere (|dipole tilt| < 10°). The figure is in the same format as Figure 3.

For all convection patterns presented in this section, the 60°–70° latitude interval is used to give a quantification of the return flow strength, on the dawn and dusk cells separately. We emphasize that the quantification of the dawn/dusk convection speed is not straightforward. Several different approaches have been tested to target the analysis on the return flow only. We have looked at the latitudinal profile of the convection, but distinguishing flow on open and closed field lines from such a profile is not trivial, as the flow shear across this boundary can be very small or absent on average, especially during northward IMF and summer conditions. We therefore use 70° MLAT as our boundary, based on the convection maps presented in Figures 3–5 and their corresponding maps of the standard deviation, as seen in an example in Figure 2c. Although the ratios change slightly depending on the method used, the overall trends still remain. We therefore consider our observational results as robust.

### 3.2. Northern Hemisphere Equinox

Figure 4 is in the same format as Figure 3 but shows observations only when the dipole tilt angle is between −10° and +10° corresponding to equinox conditions. A similar trend is seen, namely that the dawn/dusk convection ratio (the number at the bottom of each panel) approaches similar values in the bottom row (seen most clearly for southward IMF). This means that the asymmetry between positive and negative IMF $B_y$ as seen in the top row decreases with increasing tail activity (AL). The asymmetries are not as strong as during the local winter conditions and appear to mostly vanish when AL< −50 nT. However, when IMF is northward, the most active AL plots, Figures 4i and 4j, indicate the appearance of asymmetry again, deviating from the trend mentioned above.

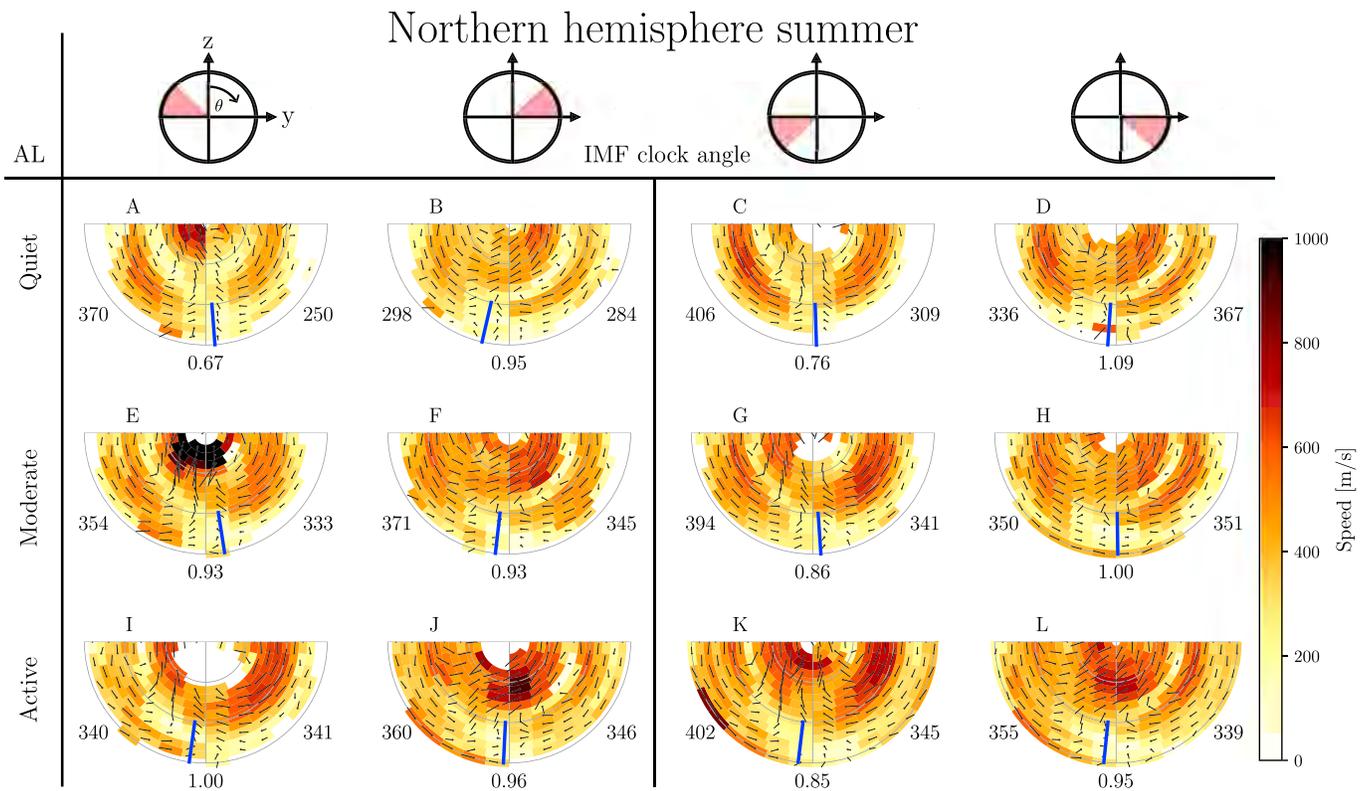

**Figure 5.** Maps of nightside ionospheric convection during summer conditions in Northern Hemisphere (dipole tilt > 10°). The figure is in the same format as Figure 3.

### 3.3. Northern Hemisphere Summer

Local summer plots, in the same format, are shown in Figure 5. A similar trend is seen with respect to decreasing asymmetries when |AL| increases, leaving the ratio values similar in the bottom row compared to the asymmetric values between Figures 5a and 5b, and 5c and 5d in the top row. The southward IMF panels show a more systematic decrease of asymmetry compared to the northward IMF panels; however, the maps during northward IMF are based on less data, as can be seen by the number of white grid cells. We also notice that the summer plots in Figure 5 indicate more plasma circulation inside the polar cap compared to the winter maps in Figure 3. This is likely related to increased lobe reconnection during local summer (Crooker & Rich, 1993).

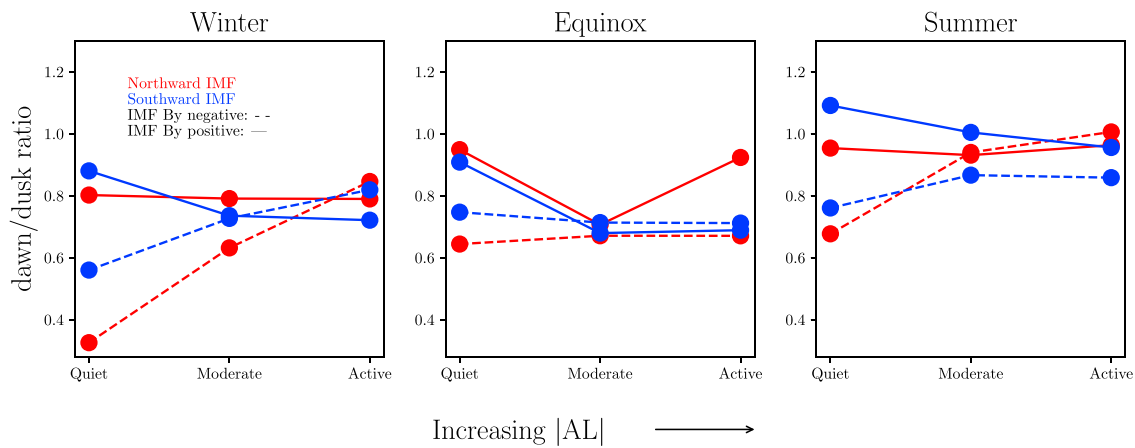

**Figure 6.** Summary of the dawn/dusk convection speed ratio from Figures 3 to 5. Winter, equinox, and summer conditions are shown separately in the three panels. Red lines are used for the northward IMF clock angles, and blue lines for the southward IMF conditions. Dashed and solid lines are used for the negative and positive IMF $B_y$ conditions, respectively. The mentioned trend can be seen by comparing lines with same the color. IMF = interplanetary magnetic field.

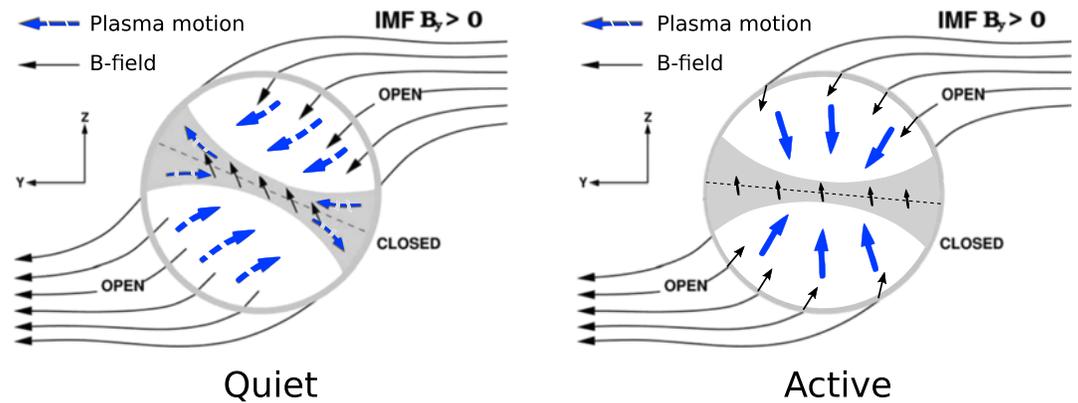

**Figure 7.** Cross sections of the magnetosphere seen from the tail toward the Earth during periods of IMF $B_y > 0$ conditions, adopted from Liou and Newell (2010). Closed field line region is indicated with gray, blue dashed arrows illustrate plasma motion, and magnetic field lines are shown as solid black lines. (left): No tail reconnection. Plasma motion is asymmetric in both lobes and closed field line region. The latter represents a shear flow that induce $B_y$ on closed field lines. The flows equilibrate the magnetic energy density asymmetry between dawn and dusk. (right): Tail reconnection is significant. The plasma motion in the lobes now converge toward the tail reconnection region in the center. The convergence of the flow reduces $\frac{dv_y}{dz}$ and hence the induced $B_y$. The resulting induced $B_y$ on the closed field lines is illustrated with the black lines inside the gray region. IMF = interplanetary magnetic field.

## 4. Discussion

Due to the large-scale Parker spiral structure of the IMF, asymmetric forcing of the magnetosphere is rather the norm than the exception. It is of interest to describe how the asymmetric state of the system behaves in response to energy conversion inside the magnetosphere (tail reconnection) and not only describe the asymmetric state as a static property that is only dependent on the SW/IMF interactions.

Our main point from our observational data presented in Figures 3–5 is more dawn/dusk asymmetric convection on closed field lines during low levels of tail activity compared to higher levels of tail activity. To highlight this result, we have plotted the trend of the convection speed ratio with |AL| in Figure 6. Winter, equinox, and summer are here shown separately in the three panels, representing three separate tests. The red lines refer to the northward IMF maps in Figures 3–5, while blue lines correspond to the southward IMF convection maps. Furthermore, we use dashed and solid lines to distinguish the negative and positive IMF $B_y$ conditions, respectively. The AL intervals on the x axis correspond to those used in Figures 3–5 and referred to as quiet, moderate, and active conditions. From Figure 6 it is easier to see how the dawn/dusk convection speed ratio becomes more similar between the dashed and solid lines as tail activity increases. Looking at the difference between the solid and dashed lines (comparing positive and negative clock angles), the difference is less during the active periods compared to the quiet periods for all seasons and both during northward and southward IMF. This opposite trend of the dashed and solid lines indicates more symmetric convection during active times and is the main observational result from this study. We also mention that there is a local minimum in the solid red curves during equinox and summer for moderate activity.

In this paper we argue that low levels of tail activity are favorable for an asymmetric closed magnetosphere by using arguments based on the observational results from this study and the conceptual understanding of how $B_y$ is induced as suggested by Khurana et al. (1996) and Tenfjord et al. (2015, 2017, 2018). In this view there are two competing processes determining the degree of asymmetry, related to the large-scale forcing of the system (Cowley & Lockwood, 1992): (1) asymmetric loading/rearranging of magnetic energy density in the lobes from dayside interactions due to IMF $B_y$ and (2) tail reconnection that annihilate magnetic energy, creating a region of decreased magnetic energy density (pressure) affecting the plasma flow in the two lobes in a more symmetric manner. Hence, the plasma flow in the two lobes will be directed toward the reconnection site, decreasing the shear flows that induce $B_y$ initiated by process 1. A simple sketch of how we anticipate the plasma motion to be affected by tail reconnection is shown in Figure 7. The left panel in Figure 7 is adopted from Figure 3a in Liou and Newell (2010) and represents the situation when mainly dayside loading/rearranging of flux is present. The right panel in Figure 7 is a modification of the same figure from Liou and Newell (2010) and shows the situation when also significant tail reconnection is ongoing.

The sketch illustrates IMF $B_y$ positive conditions, and the perspective is looking toward the nightside Earth from the tail. Blue dashed arrows indicate plasma motion, and solid black lines indicate the magnetic field in both the open (white) and closed (gray) field line regions. According to Tenfjord et al. (2015) equation (4), $\frac{dv_y}{dz}$ is contributing to the induced $B_y$ on closed field lines. In our sketch in Figure 7 we try to emphasize this, where the quiet (left) panel shows a large $\frac{dv_y}{dz}$ on closed field lines allowing $B_y$ to develop. The $\frac{dv_y}{dz}$ arises as plasma in the northern/southern parts of the closed field lines move duskward/dawnward, respectively, in response to the asymmetric pressure distribution that has been introduced. The active (right) panel shows a more converging flow toward the reconnection site, which act to fill the void left by the closure of flux. If the pressure distribution is mainly determined by the reconnection process in the active (right) panel, the asymmetric pressure contribution from the lobes will be of minor importance and a $\frac{dv_y}{dz}$ on closed field lines is not expected. We therefore propose a more converging flow toward the reconnection site suggesting that tail reconnection could limit the induced $B_y$ due to the asymmetric loading or rearranging of magnetic flux. The effect on the closed field lines is illustrated with the black lines in the gray region. We emphasize that the the purpose of the blue arrows is to represent the direction, not magnitude, of how the asymmetric loading and tail reconnection can affect the plasma motion locally.

This understanding has important consequences for the process of restoring symmetry, which is when one end of an asymmetric field line gradually catches up with the other hemisphere as it convects toward the dayside via dawn or dusk (Reistad et al., 2016). As argued above, tail reconnection will make the magnetic energy density distribution in the tail more symmetric, leading to more symmetric closed field lines as they reconfigure to the new situation. Hence, there will be less asymmetry that needs to be restored when field lines convect toward the dayside via the dawn or dusk convection cell. This provides an explanation for the asymmetric TRINNI flows being most frequently seen during northward, but $B_y$-dominated periods (Grocott et al., 2008), as these are the conditions when tail activity is low and the field can become highly asymmetric.

There will also be an additional effect further enhancing the rate at which the nightside closed field lines become symmetric. For low levels of tail activity, magnetic flux transport from the tail toward the dayside is small. Hence, the closed field lines in the nightside have had a long time to reconfigure in response to the asymmetric loading/rearranging of magnetic flux, and the closed magnetosphere can become highly asymmetric. When tail activity increases, the asymmetric field itself will start to convect toward the Earth where the restoring of symmetry takes place. After a period of enhanced tail reconnection, the nightside closed field lines will mostly consist of recently closed field lines having no or only a small $B_y$ component due to the arguments mentioned above. Hence, reconnection also acts to evacuate the asymmetric field itself.

The net result of these processes is that during active times, the nightside closed field lines will become more symmetric, as indicated by the black lines in Figure 7. In the steady state, corresponding to the results presented in the previous section, the process of restoring symmetry does not take place or is less pronounced during geomagnetic active conditions, as there is no or only little asymmetry to restore. Our observations are consistent with this interpretation, so is also the study by Grocott et al. (2010), looking at the global convection in response to substorms for different IMF $B_y$ conditions. Although they focused on influences on the convection *pattern* rather than convection *speed* as we do in this study, they found the substorm to produce a similar pattern in the nightside closed auroral region, independent of IMF $B_y$. This is also evident in our plots from the blue line separating the dawn and dusk cells in the nightside. In the bottom row panels i–l (active times) in Figures 3–5, this location is similar for all IMF clock angles, whereas in the corresponding top row panels a–d (quiet times) the location of the blue line varies much for the different IMF clock angle sectors. This we also interpret as the magnetotail being more symmetric during the more geomagnetic active conditions.

Tenfjord et al. (2015) presented a theoretical framework for how IMF $B_y$ influences (during southward IMF) the closed magnetosphere by considering the forces that lead to asymmetric footprints on closed magnetic field lines. In that description, it is the forces due to an asymmetric magnetic energy density distribution (a result of the asymmetric loading of flux from the dayside) that make the closed field lines, sandwiched between the lobes, to move in opposite directions in each hemisphere. This is what lead to an induced $B_y$ component on closed field lines and asymmetric footprints in this description. The conceptual understanding outlined above has been further supported by later studies (Tenfjord et al., 2017, 2018) looking at the time needed to see the effect of an induced $B_y$ at geostationary orbit. Since the magnetic energy density starts to become asymmetric only minutes after the start of asymmetric loading, the response in $B_y$ at geostationary orbit was consistently found to be visible at that timescale. These results clearly demonstrate that tail reconnection of open field lines is not necessary to produce closed field lines with a $B_y$ component and indicate that the magnetosphere almost immediately starts to adjust to the the present pressure distribution, with no sharp distinction between the open and closed field lines. This finding is somewhat in contrast to other studies on this topic. Cowley (1981) and Fear and Milan (2012) suggest that asymmetric footprints and $B_y$ are first introduced into the closed magnetosphere when lobe flux closes to form the plasma sheet, implying a significantly longer response time than the one found by Tenfjord et al. (2017, 2018). Browett et al. (2017) looked at the plasma sheet $B_y$ in response to IMF $B_y$ using the Cluster spacecraft and found timescales more consistent with the Cowley (1981) and Cowley and Lockwood (1992) description. Clearly, how and where $B_y$ enter the magnetosphere for various SW/IMF and tail reconnection levels need further investigations. The present study does not address the timescales at which the closed magnetosphere become more symmetric. We can therefore not conclude from our data whether tail reconnection is responsible for further enhancing $B_y$ in the closed magnetosphere on top of what is induced due to the asymmetric pressure. This remains to be investigated in more detail.

The arguments presented above regarding how increased tail reconnection acts to reduce asymmetries in the closed magnetosphere are also supported by earlier interhemispheric observations of the aurora from space through the different substorm phases (Østgaard et al., 2011). During a 5-hr interval with fairly stable IMF and including two substorm onsets, they were able to determine the displacement of conjugate auroral features in the nightside. Their results showed convincingly that the longitudinal displacement of conjugate regions, most significantly the substorm onset region, decreased through the expansion phase of each substorm. Another relevant study in this regard is the investigation of magnetic field perturbations at geosynchronous orbit during various levels of geomagnetic activity and IMF clock angle (Cowley & Hughes, 1983). They show in their Figure 2 that the disturbance in $B_y$ at geostationary orbit is positively correlated with IMF $B_y$, but the disturbance decreases with increasing $K_p$ index. However, no explanation was provided.

Based on the arguments and observations presented in this study, we would also expect to observe a dawn/dusk asymmetry in the convection speed on closed field lines at the initial stages of a substorm, if the external forcing had been asymmetric prior to the substorm. An additional effect during substorms is that the magnetosphere inflates (deflates) during the growth (expansion) phase when flux is opened (closed). This will lead to an increase (decrease) of the flaring of the magnetosphere and hence the pressure in the lobes (Caan et al., 1975), possibly enhancing the efficiency of the asymmetric loading in the growth phase, and the rate at which the pressure becomes more symmetric in the expansion phase. Provan et al. (2004) did a superposed epoch analysis of ionospheric convection derived from SuperDARN during 67 substorm intervals in the Northern Hemisphere, with emphasis on the convection speed. In their Figure 15 they show the average high-latitude ionospheric convection pattern at substorm onset during IMF $B_y$ positive- and negative-dominated substorms, separately. Provan et al. (2004) noted that stronger average convection was consistently observed on the banana-shaped convection cell, namely in the dusk cell during negative IMF $B_y$ and in the dawn cell during positive IMF $B_y$. However, they did not provide any information regarding how this asymmetry evolved with time during the expansion phase, as their superposed epoch analysis was performed for all IMF $B_y$. The present study suggests that the substorm will act to reduce such asymmetries present in an asymmetric tail. However, this remains to be investigated in more detail.

## 5. Conclusion

The main findings from this study can be summarized in the following points:

1. The level of tail activity affects the relative convection speed toward the dayside between the dawn and dusk convection cells in the same hemisphere during periods of asymmetric loading/rearranging of magnetic flux.
2. Tail reconnection and hence substorms act to reduce the asymmetric state of the closed magnetosphere by reducing the asymmetric pressure introduced from IMF $B_y$ interacting with the magnetosphere.
3. In addition to the convection speed on closed field lines, also the nightside convection pattern is observed to become more symmetric for increasing levels of tail activity.

**Acknowledgments**
SuperDARN (Super Dual Auroral Radar Network) is an international collaboration involving more than 30 low-power HF radars that are operated and funded by universities and research organizations in Australia, Canada, China, France, Italy, Japan, Norway, South Africa, United Kingdom, and USA. The convection data were retrieved as "gridex files" from Virginia Tech using the DaViTpy software (https://github.com/vtsuperdarn/davitpy). We acknowledge the use of NASA/GSFC's Space Physics Data Facility (http://omniweb. gsfc.nasa.gov) for OMNI data. Kjellmar Oksavik is grateful for being selected as the 2017–2018 Fulbright Arctic Chair, and his sabbatical at Virginia Tech is sponsored by the U.S.-Norway Fulbright Foundation for Educational Exchange. Financial support has also been provided to the authors by the Research Council of Norway under the contract 223252. A. G. is supported by STFC grant ST/M001059/1 and NERC grant NE/P001556/1.